\documentclass[12pt]{JHEP}
\usepackage{epsfig}

\def\la{\mathrel{\mathpalette\fun <}}
\def\ga{\mathrel{\mathpalette\fun >}}
\def\fun#1#2{\lower3.6pt\vbox{\baselineskip0pt\lineskip.9pt
  \ialign{$\mathsurround=0pt#1\hfil##\hfil$\crcr#2\crcr\sim\crcr}}}
\newcommand{\beq}{\begin{equation}}
\newcommand{\eeq}{\end{equation}}
\title{On the propagation of the highest energy cosmic ray nuclei}

\author{Luis N. Epele and Esteban Roulet\\Depto. de F\'\i sica, 
Universidad Nacional de La Plata\\ CC67,  1900, La Plata, Argentina
\\ \email{roulet@venus.fisica.unlp.edu.ar}}

\abstract{We study the propagation of ultra-high energy cosmic ray 
nuclei through the background of cosmic microwave and intergalactic 
infrared photons, using recent re-estimates for the density of the
last ones. We perform a detailed Monte Carlo simulation to follow 
the disintegration histories of nuclei starting as Fe and 
reaching the Earth from extragalactic sources.
We obtain the maximum energies of the arriving nuclear fragments as
well as the mass composition as a function of the distance traveled.
Cosmic rays with energies in excess of $2\times 10^{20}$~eV cannot 
originate from Fe nuclei produced in sources beyond 10~Mpc.}


\begin{document}

\section{Introduction}

The cosmic ray (CR) spectrum is known to extend up to energies 
beyond $10^{20}$~eV, with the highest energy air showers observed 
having energies of 2--3$\times 10^{20}$~eV. The origin and nature 
of these ultra-high energy (UHE) events are one of the pressing unsolved 
problems defying us today. The CR spectrum has the overall shape 
of a leg, and is well fitted by power laws, whose index increases
(spectrum steepening) for energies
 above the `knee' ($E\sim 3\times 10^{15}$~eV),
flattening again above the `ankle' (at $E\sim 5\times 10^{18}$~eV).
The CR composition becomes heavier for increasing energies around 
the knee, and the CR are probably mostly of galactic origin up to the
ankle. Approaching the ankle,  the CR composition 
 seems to become lighter again, and  the increasing rigidity of CRs does 
not allow anymore their confinement into the Galaxy. Hence, CR fluxes are 
most probably of extragalactic origin above the ankle.  
 There have been studies suggesting
that the arrival direction of the highest energy events may be indicating
that their origin lies in the local supercluster, but they are not
conclusive. The small anisotropies observed may also be compatible
with a cosmological origin of the highest energy events.

The big difficulty which appears is that CR protons with $E\ga 5\times 
10^{19}$~eV, 
i.e. relativistic $\gamma$ factors $\ga 5\times 10^{10}$, are not able to 
propagate more than $\sim 100$~Mpc due to their energy losses by
photopion production off the cosmic microwave background (CMB)
 photons, giving rise to the well known GZK cutoff \cite{gzk}. At 
energies $2\times 10^{20}$~eV their mean free path is already only 30~Mpc.
Heavy nuclei with smaller $\gamma$ factors, but comparable energies, 
also get attenuated but mainly by photodisintegrations off the 
intergalactic infrared (IR) background and off 
 CMB photons, as well as by pair creation losses to a lesser extent
\cite{st69,tk75,pu76,el95}.

A detailed study of the propagation of UHECR Fe nuclei, including all the 
relevant energy loss mechanisms, was performed more than twenty years
ago by Puget, Stecker and Bredekamp \cite{pu76}. However, the 
estimates of the density of IR photons employed then were about 
an order of magnitude larger than the new empirically based
estimates obtained 
using the measured emissivity of IRAS galaxies \cite{ma98}.
In the light of the lower IR background densities inferred recently, 
it was suggested  that UHECR nuclei may propagate much longer
distances unattenuated \cite{st98}, so that the events with energies
2--3$\times 10^{20}$~eV could have possibly originated as Fe nuclei
produced at distances up to 100~Mpc, and hence in particular 
in the whole local supercluster\footnote{It has to be
stressed that Fe nuclei are good candidates for UHECRs, due to
their high abundance in supernova environments and their
large value of $Z$, which enhances the energy achievable in 
the acceleration process.}.  However, as we showed in a 
recent letter \cite{ep98}, at energies larger than $10^{20}$~eV it
is photodisintegration off CMB photons (and not off IR ones) which
dominates the opacity for Fe disintegration. This implies
 that the maximum
energies with which the surviving fragments can reach the Earth are 
not significantly changed   (for distances below a few hundred Mpc) 
with the new estimates of the IR density. In particular, for 
sources at distances of  100~Mpc the maximum energies of the 
surviving fragments do not exceed $\sim 10^{20}$~eV, and can arrive
to $2\times 10^{20}$~eV only for distances below 10~Mpc.

The aim of the present paper is to re-evaluate in detail the
propagation of heavy nuclei, following the photodisintegration histories
by means of a Monte Carlo which includes all relevant processes, 
much in the spirit of the original Puget et al. paper \cite{pu76}.
From this we can establish all the effects resulting from the new 
estimates of the IR background density. In particular, we obtain 
the final mass composition and energy as a function of the distance to
the source, as well as the possible fluctuations in these 
quantities which may arise from the particular way in which
the photodisintegration takes place in each case. We also
study the effects of pair creation losses, which turn out to be 
important in some cases for the determination of the final mass
composition.

\section{The propagation of heavy nuclei}

As we said before, CR with energies above the ankle are most
probably extragalactic. This means that in their journey they 
may be attenuated by the interactions with the photon background.
This background consists essentially of the microwave photons
of the 2.7$^\circ$K cosmic background radiation and, at larger
energies, of the intergalactic background of IR photons emitted
by galaxies. The background of optical radiation turns out to 
be of no relevance for UHECR propagation.

Although the CMB density is well known, the intergalactic IR one
cannot be measured directly and has to be estimated from the observation
of the spatial distribution, IR spectra and emissivity of the
galaxies which are sources for this IR emission.
This was done recently by Malkan and Stecker \cite{ma98}, 
who obtained a result which is about an order of magnitude smaller than 
previous estimates. We will then adopt for the spectral density
of the IR background
\begin{equation}
{{\rm d}n\over {\rm d}\epsilon}=1.1\times 10^{-4}\left({\epsilon\over {\rm eV}}
\right)^{-2.5}\ {\rm cm^{-3}eV^{-1}}
\end{equation}
for photon energies $\epsilon$ in the range between $2\times 10^{-3}$~eV
and 0.8~eV. This is a factor of 10 smaller than the ``high infrared 
(HIR)'' density adopted in ref.~\cite{pu76}, and is in the upper
range of the recent estimates.
In order to quantify the possible effects of an optical intergalactic
background, we just modeled this last with a Planckian 
distribution with $T=5000^\circ$K and a dilution factor 
of $1.2\times 10^{-15}$, as in 
\cite{pu76}.

UHECR nuclei propagating through these photon backgrounds will
loose energy mainly by two processes:

$i)$ photopair production, which has a threshold corresponding to 
photon energies in the rest frame of the nucleus of $2m_ec^2
\simeq 1$~MeV. 
This 
process was studied in detail by Blumenthal \cite{bl70}, and 
the main contribution arises from interactions with CMB
photons. We used for the energy loss rate the expressions given 
in ref.~\cite{ch92}.

$ii)$ photodisintegration losses, for which the rate of emission 
of $i$ nucleons from a nucleus of mass $A$  (with cross section 
$\sigma_{A,i}$) is given by
\begin{equation}
R_{A,i}={1\over 2\gamma^2}\int_0^\infty {{\rm d}\epsilon \over
\epsilon^2}{{\rm d}n\over {\rm d}\epsilon}\int_0^{2\gamma \epsilon}
{\rm d}\epsilon' \epsilon'\sigma_{A,i}(\epsilon'),
\end{equation}
where $\gamma$ is the relativistic factor of the nucleus, 
$\epsilon$ the photon energy in the observer's system and $\epsilon'$
its energy in the rest frame of the CR nucleus.

The cross sections for photodisintegration $\sigma_{A,i}(\epsilon')$
contain essentially two regimes. At $\epsilon'<30$~MeV there is the 
domain of the giant resonance and the disintegration proceeds mainly
by the emission of one or two nucleons. At higher energies, the cross 
section is dominated by multi-nucleon emission for heavy nuclei 
and is approximately flat up to $\epsilon'\sim 150$~MeV. We fitted the 
various $\sigma_{A,i}$ with the parameters in Table I and II of 
ref.~\cite{pu76}.  A useful quantity to estimate the energy loss rate
by photodisintegration is given by the effective rate
\begin{equation}
R_{eff,A}={{\rm d}A\over {\rm d}t}=\sum_i iR_{A,i}.
\end{equation}

\EPSFIGURE{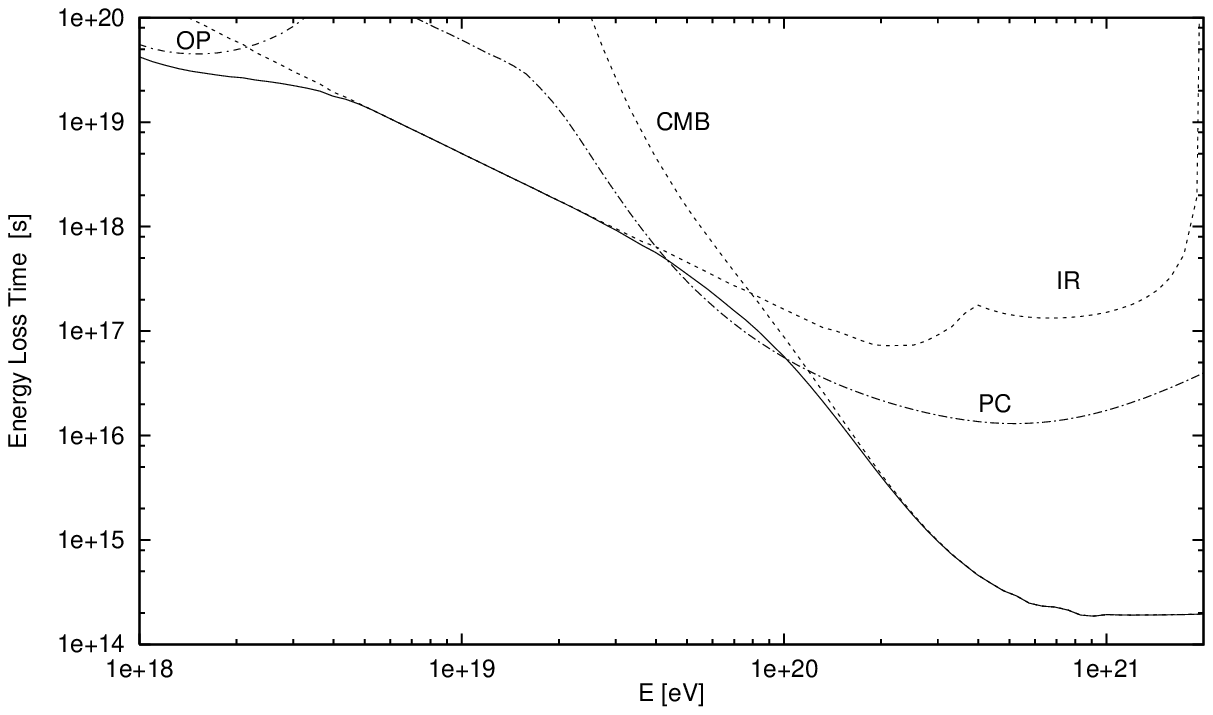}{Effective energy loss time for Fe
photodisintegration
off microwave (CMB), infrared (IR) and optical (OP) photons, as 
well as the total one (solid line) and the pair creation (PC) 
energy loss time.}

Since neglecting pair creation processes one has that 
 photodisintegrations alone lead to
 $E^{-1}$d$E/$d$t=A^{-1}$d$A/$d$t$, 
the energy loss time for photodisintegration is then $A/R_{eff,A}$.
The different contributions to this quantity 
 are plotted in Fig.~1. We show separately the contributions 
to the disintegration from CMB, IR and 
optical photons for Fe nuclei,  together with the total 
 one (solid line) and the photopair creation energy 
loss rate\footnote{Looking at this and the following figures, 
it is important to keep in mind that 1~Mpc$=1.03\times 10^{14}$~s.}.

It is apparent that the optical background has no relevant effect, 
that the IR one dominates the photodisintegration processes below 
$10^{20}$~eV and the CMB dominates above $10^{20}$~eV. The pair 
creation rate is relevant for Fe energies $4\times 10^{19}$~eV--$2\times 
10^{20}$~eV (i.e. $\gamma$ factors $\sim 1$--$4\times 10^9$), for
which the typical CMB 
photon energy in the rest frame of the nucleus is above 
threshold ($>1$~MeV) but still well below the peak of the giant resonance
($\sim 10$--20~MeV). The effect of pair creation losses is to reduce
the $\gamma$ factor of the nucleus, obviously leaving $A$ unchanged.

\section{Results}

Using the rates just discussed, we performed a Monte Carlo
simulation in order to follow the possible disintegration 
histories of Fe nuclei. In figures 2 and 3 we plot the final 
mass (i.e. $A$) and energy $E$ of the heaviest fragment surviving 
from the disintegration process. We show in these figures the results
of simulations with initial values of the relativistic factor 
 $\gamma_0=1\times 10^{10}$, 
$4\times 10^9$ and $1\times 10^9$. The curves shown for each 
value of $\gamma_0$ correspond, in Fig.~2, to the average value
$\langle A\rangle$ from all the simulations (solid line) and
 the region (between 
the two dashed lines) including 95\%  of the simulations\footnote{
Only 2.5\% of the simulations are below the lower curves and 2.5\% 
are above the upper ones.}. This gives a clear idea of the range of
values which can result from fluctuations from the average behaviour.

\EPSFIGURE{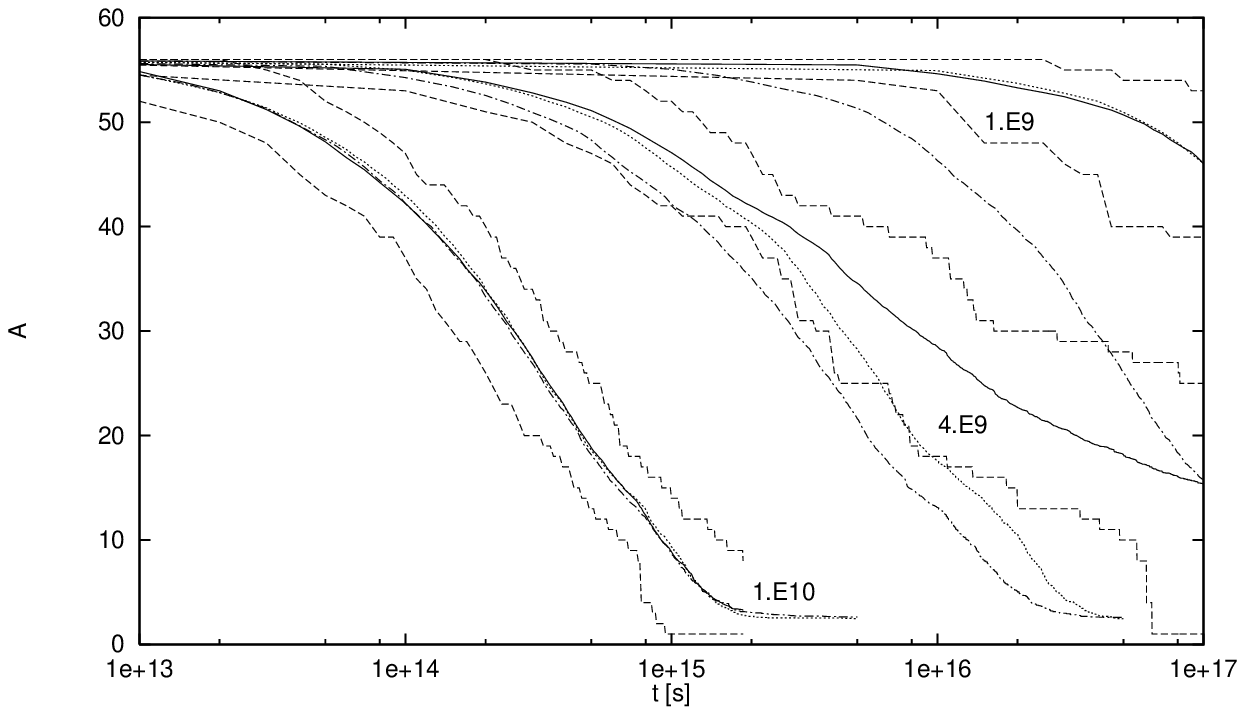}{Evolution of the mass number $A$ of the 
heaviest fragment surviving photodisintegration vs. travel time $t$.
The initial $\gamma$ factors considered are $1\times 10^{10}$ 
(1.E10 curve), $4\times 10^9$ (4.E9 curve) and $1\times 10^9$ (1.E9).}

To further understand the relevance of the different processes 
and the impact of the new determinations of the IR density, we also plot
 the results for $\langle A\rangle$ obtained in simulations 
which do not include pair creation processes (dotted lines) and also the
results we would obtain  (dot-dashed line) with an IR density a factor
of ten larger (i.e. the HIR density of ref.~\cite{pu76}).
Figure~3 is similar but for the final values of the energy. 

\EPSFIGURE{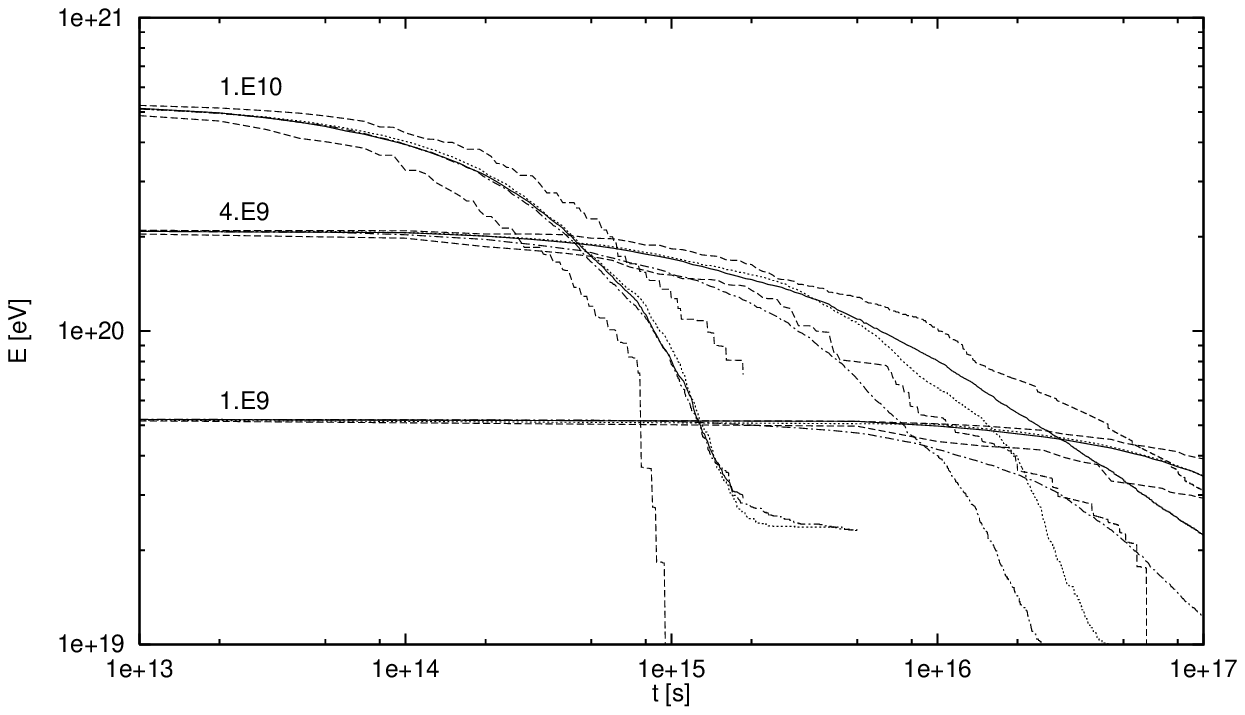}{Similar as Figure 2 but for the final energy 
$E$ vs. $t$.}

From these figures we can draw the following conclusions:

$i)$ For large initial energies ($E_0>2.5\times 10^{20}$~eV, i.e. 
$\gamma_0>5\times 10^9$), both the effects of the IR photons and of pair
creation processes are of no relevance along the whole
journey of the nucleus, and the energy losses are 
essentially due to photodisintegration off CMB photons alone.

$ii)$ At $\gamma_0<5\times 10^9$ the pair creation losses start 
to be relevant, reducing the value of $\gamma$ significantly 
as the nucleus propagates distances $O$(100~Mpc).
The effect is maximum for $\gamma_0\simeq 4\times 10^9$ but becomes 
small again for $\gamma_0\la 1\times 10^9$, for which appreciable effects 
only appear for cosmological distances ($>10^3$~Mpc),
as can be simply understood
from Fig.~1. 
The effect of neglecting pair creation losses translates into keeping
$\gamma=\gamma_0$ constant during the propagation, and this enhances the
photodisintegration rates and then reduces $\langle A\rangle$
 more rapidly.

$iii)$ Also for $\gamma_0<5\times 10^9$ the reduction in the IR
density adopted has sizeable effects. In this respect point $ii)$ is 
relevant, since pair creation losses shift the values of $\gamma$ 
towards a domain where IR photons become increasingly important with
respect to CMB ones. With the new values of the IR density the 
effects of photodisintegrations become small already for 
$\gamma_0\simeq 1\times 10^9$ if we consider
 propagation distances below $10^3$~Mpc
(i.e. for $t<10^{17}$~s).

$iv)$ The effects of neglecting pair creation losses are less 
pronounced in Fig.~3. For instance, for $\gamma_0=4\times 10^9$ 
the average energies with and without pair creation processes
are similar up to $t\simeq 10^{16}$~s while the $\langle A\rangle$
values differ sizeably already for $t\simeq 3\times 10^{15}$~s.
This is due to a partial cancellation between the 
effects of the evolution of $\gamma$ and of $A$ in the values of the
final energy ($E=m_p\gamma A$), since neglecting pair creation losses
does not allow $\gamma$ to decrease but makes instead $A$ to drop
faster\footnote{This in particular shows that the 
inclusion of pair creation losses does not modify 
the maximum attainable energies
computed in \cite{ep98}.}.

$v)$ The effects of fluctuations due to different photodisintegration
histories are not negligible. They give a spread in $A$ (and $E$) 
 of the order of 10\% (considering the 95\% probability range) 
for $\langle A\rangle \simeq 40$ 
but relatively larger for smaller $\langle A\rangle$, since variations
$\Delta A\sim 10$--15 at a given time $t$ can appear between 
different simulations.

\EPSFIGURE{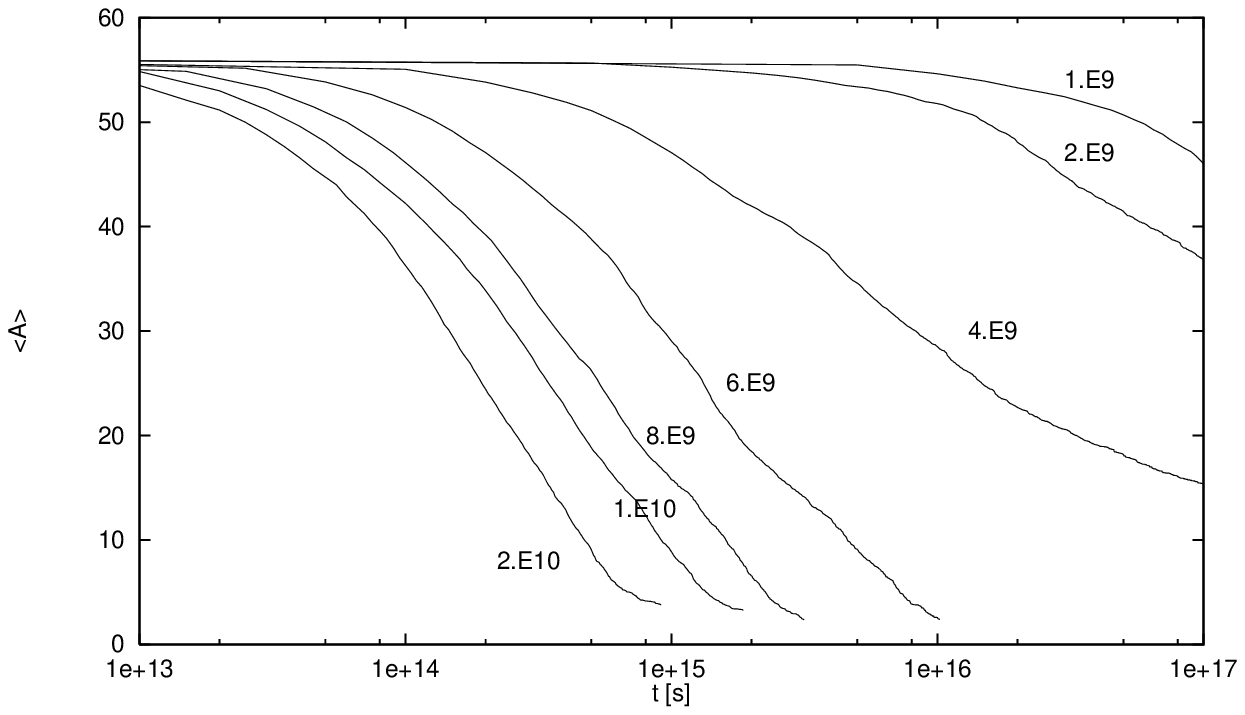}{$\langle A\rangle$ vs. $t$ for more sample 
values of $\gamma_0$.}

In figures 4 and 5 we plot more sample values of $\langle A\rangle$ and 
$\langle E\rangle$, for values of $\gamma_0=2\times 10^{10}$, $1
\times 10^{10}$ and (8, 6, 4, 2, 1)$\times 10^9$. Looking at Fig.~5
it is easy to infer the maximum energies which can be obtained
from Fe nuclei injected at any fixed distance $d$. 
In particular, for $d=100$~Mpc ($t=10^{16}$~s) the maximum average 
energy is $E_{max}\simeq 8\times 10^{19}$~eV and originates from
$\gamma_0\simeq 2$--4$\times 10^9$. Comparing with Fig.~4 we see that
these maximum energy events would correspond to fragments with
masses $A(E_{max})\simeq 30$--50, i.e. a rather heavy composition.

Fluctuations from the average behaviour can only slightly 
increase the maximum  attainable energies, and this is illustrated 
with the dashed line, which represents the upper boundary of the 
95\% CL ranges (i.e. 97.5\% of the simulations are below this curve)
for any initial value of $\gamma_0$.

For source distances $d= 10$~Mpc, average energies up to
$E_{max}\simeq 2\times 10^{20}$~eV can result, and an interesting
pile-up effect is observed since a broad range of initial energies
(with $\gamma_0\sim 4$--$8\times 10^9$) lead to approximately the same
final energy ($\sim E_{max}$). This can produce a bump in the spectrum
from sources at these distances if indeed the highest energy events
originate from heavy nuclei.   Due to the spread in values of $\gamma_0$
at $E_{max}$, we see from a comparison
 with Fig.~4 that there will also be a 
wide spread in the final composition, with $A(E_{max})\simeq 10$--45.
Events with energies  2--$3\times 10^{20}$~eV may appear as low
probability fluctuations from the mean behaviour if $d\simeq 5$--8,
 having initially
$\gamma_0>10^{10}$ and a low mass final composition ($A<10$).

\EPSFIGURE{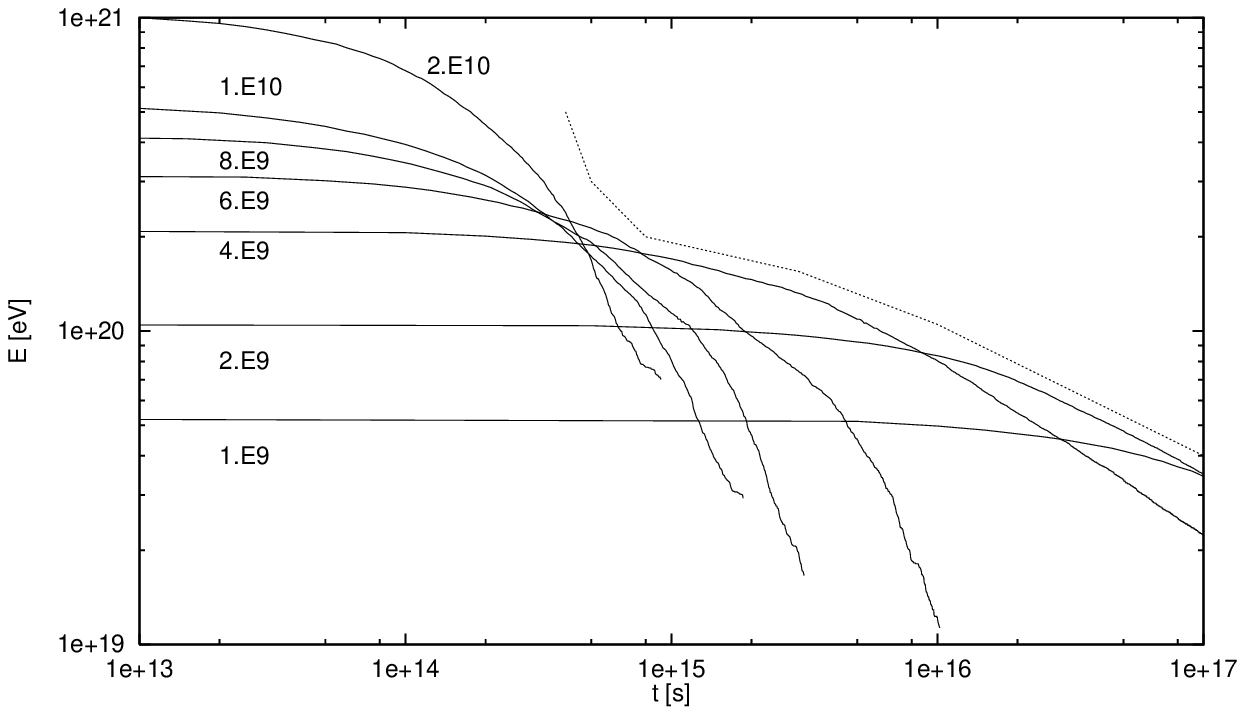}{$\langle E\rangle$ vs. $t$ for several 
sample values of $\gamma_0$. The dotted line indicates the upper 
boundaries of the 95\% probability range obtained in the simulations.}

For smaller source distances ($d\sim$ few Mpc), events with $E\simeq 
3\times 10^{20}$~eV could be heavier nuclei with smaller initial
energies ($\gamma_0>6\times 10^9$).

For very large values of $\gamma_0$ ($\gamma_0>2\times 10^{10}$), the 
heavy nuclei completely disintegrate in less than 10~Mpc, and the 
photopion production (not included here) becomes the main 
attenuation process for the secondary nucleons, which are then 
subject to the usual GZK cutoff.

In conclusion, the main implication of the lower values of the 
IR density recently estimated is to increase the mean free path 
of the heavy nuclei with initial 
$\gamma$ factors below $\sim 5\times 10^9$, 
for which most CMB  photons are below the peak of the giant resonance
for photodisintegration.
Due to the fragmentation of the nuclei by photodisintegration 
and the pair creation energy losses, the final energies of the 
fragments are typically below $2\times 10^{20}$~eV for travel distances
$\sim 10$~Mpc, and below $10^{20}$~eV for distances $\sim 100$~Mpc.
The new value of the IR density is then of little help in the attempts 
to understand the highest energy events observed ($E\sim 2$--$3\times 
10^{20}$~eV), which could not have originated as heavy nuclei at
distances beyond $\sim 10$~Mpc. The lack of obvious candidate sources
at closer distances \cite{el95} 
leave the nature and origin of these events still 
 a mistery.

\acknowledgments

Work partially supported by CONICET, Argentina.
We thank Luis Anchordoqui and M. Teresa Dova for discussions.

\end{document}